\documentclass[preprint2,numberedappendix]{emulateapj-rtx4}
\usepackage{graphicx,natbib,bm,url,color,mathtools}

\graphicspath{{./fig/}{./png/}}
\newcommand{\EQ}{\begin{equation}}
\newcommand{\EN}{\end{equation}}
\newcommand{\EQA}{\begin{eqnarray}}
\newcommand{\ENA}{\end{eqnarray}}

\newcommand{\EEq}[1]{Equation~(\ref{#1})}
\newcommand{\Eq}[1]{Equation~(\ref{#1})}

\newcommand{\App}[1]{Appendix~\ref{#1}}

\newcommand{\Sec}[1]{Sect.~\ref{#1}}

\newcommand{\Fig}[1]{Figure~\ref{#1}}

\newcommand{\Figss}[2]{Figures~\ref{#1}--\ref{#2}}
\newcommand{\Tab}[1]{Table~\ref{#1}}

\newcommand{\bra}[1]{\langle #1\rangle}

{}
{}
{}

{}
{}
{}
{}
{}
{}
{}
{}
{}
{}
{}
{}
{}
{}
{}
{}
{}

{}

{}

{}
{}
{}

\newcommand{\tildeA}{\tilde{A}}
\newcommand{\tildeB}{\tilde{B}}

\newcommand{\FLASH}{{\tt FLASH}~}
\newcommand{\FLASHz}{{\tt FLASH}}

\newcommand{\meanAA}{{\overline{\bm{A}}}}
\newcommand{\meanBB}{{\overline{\bm{B}}}}
\newcommand{\meanJJ}{{\overline{\bm{J}}}}

\newcommand{\aaaa}{\bm{a}}
\newcommand{\jj}{\bm{j}}

\newcommand{\bb}{\bm{b}}
\newcommand{\BB}{\bm{B}}

\newcommand{\JJ}{\bm{J}}

\newcommand{\AAA}{\bm{A}}

\newcommand{\uu}{\bm{u}}

\newcommand{\ff}{\mbox{\boldmath $f$} {}}

\newcommand{\nab}{{\bm{\nabla}}}

\newcommand{\ii}{{\rm i}}

\newcommand{\dd}{{\rm d} {}}

\def\degr{\hbox{$^\circ$}}
\def\la{\mathrel{\mathchoice {\vcenter{\offinterlineskip\halign{\hfil
$\displaystyle##$\hfil\cr<\cr\sim\cr}}}
{\vcenter{\offinterlineskip\halign{\hfil$\textstyle##$\hfil\cr<\cr\sim\cr}}}
{\vcenter{\offinterlineskip\halign{\hfil$\scriptstyle##$\hfil\cr<\cr\sim\cr}}}
{\vcenter{\offinterlineskip\halign{\hfil$\scriptscriptstyle##$\hfil\cr<\cr\sim\cr}}}}}

\def\Pm{\mbox{\rm Pr}_{\rm M}}
\def\Rm{\mbox{\rm Re}_{\rm M}}

\def\EEK{{\cal E}_{\rm K}}
\def\EEM{{\cal E}_{\rm M}}

\def\EM{E_{\rm M}}

\def\kf{k_{\rm f}}
\def\kfeff{k_{\rm f}^{\rm eff}}

\def\HC{H_{\rm C}}
\def\HM{H_{\rm M}}
\def\EM{E_{\rm M}}

\def\urms{u_{\rm rms}}

\def\etatz{\eta_{\rm t0}}
\def\etatz{\eta_{\rm t0}}

\newcommand{\yapj}[3]{ #1, {ApJ,} {#2}, #3}
\newcommand{\yphysd}[3]{ #1, {PhysD,} {#2}, #3}

\newcommand{\yapjs}[3]{ #1, {ApJS,} {#2}, #3}

\newcommand{\yan}[3]{ #1, {AN,} {#2}, #3}

\newcommand{\ygafd}[3]{ #1, {GApFD,} {#2}, #3}

\newcommand{\yjfm}[3]{ #1, {JFM,} {#2}, #3}

\newcommand{\yprl}[3]{ #1, {PhRvL,} {#2}, #3}

\newcommand{\ymn}[3]{ #1, {MNRAS,} {#2}, #3}

\newcommand{\yprd}[3]{ #1, {PhRvD,} {#2}, #3}
\newcommand{\ypre}[3]{ #1, {PhRvE,} {#2}, #3}

\newcommand{\yjcp}[3]{ #1, {JCoPh,} {#2}, #3}
\newcommand{\yjour}[4]{ #1, {#2}, {#3}, #4}

\newcommand{\ybook}[3]{ #1, {#2} (#3)}

\begin{document}

\title{Magnetic helicity dissipation and production in an ideal MHD code}
\author{
Axel Brandenburg$^{1,2,3,4}$\thanks{E-mail:brandenb@nordita.org}
\& Evan Scannapieco$^{5}$
}

\affil{
$^1$Nordita, KTH Royal Institute of Technology and Stockholm University, Roslagstullsbacken 23, SE-10691 Stockholm, Sweden\\
$^2$Department of Astronomy, AlbaNova University Center, Stockholm University, SE-10691 Stockholm, Sweden\\
$^3$JILA and Laboratory for Atmospheric and Space Physics, University of Colorado, Boulder, CO 80303, USA\\
$^4$McWilliams Center for Cosmology \& Department of Physics, Carnegie Mellon University, Pittsburgh, PA 15213, USA\\
$^5$Arizona State University, School of Earth and Space Exploration, P.O. Box 871404, Tempe, AZ 85287, USA
}

\submitted{Astrophys. J. 889, 55 (2020)}
\date{Received 2019 October 14; revised 2019 November 27; accepted 2019 December 2; published 2020 January 24}

\begin{abstract}
We study a turbulent helical dynamo in a periodic domain by solving the
ideal magnetohydrodynamic (MHD) equations with the \FLASH code using the
divergence-cleaning eight-wave method and compare our results with direct numerical
simulations (DNS) using the {\sc Pencil Code}.
At low resolution, \FLASH reproduces the DNS results qualitatively by
developing the large-scale magnetic field expected from DNS, but at
higher resolution, no large-scale magnetic field is obtained.
In all those cases in which a large-scale magnetic field is generated,
the ideal MHD results yield too little power at small scales.
As a consequence, the small-scale current helicity is too small compared
with that of the DNS.
The resulting net current helicity has then always the wrong sign, and
its statistical average also does not approach zero at late times,
as expected from the DNS.
Our results have implications for astrophysical dynamo simulations of
stellar and galactic magnetism using ideal MHD codes.
\end{abstract}

\keywords{
dynamo --- magnetic fields --- MHD --- turbulence --- methods: numerical
}

\section{Introduction}

Astrophysical dynamos operate at large magnetic Reynolds numbers.
This means that at large and moderately large scales, the magnetic
diffusion term is negligible compared with the nonlinear terms.
However, some level of magnetic diffusion and viscosity is still needed
in numerical simulations to keep the code stable and to dissipate kinetic
and magnetic energies into thermal energy.
In numerical codes that solve the so-called ``ideal'' magnetohydrodynamic
(MHD) equations, this is accomplished by purely numerical means.
Owing to the dissipation of energy, especially in the presence
of turbulence, such codes can never be truly ideal.
Moreover, it is unclear to what extent the solutions resemble aspects
of resistive codes.
Nevertheless, throughout this paper, we continue talking about ideal
MHD equations, keeping this caveat in mind.

In spite of the comparatively small values of the magnetic diffusivity,
the process of magnetic diffusion is an essential part of any dynamo,
because the magnetic field evolution would otherwise be reversible.
This is illustrated by what is called the stretch--twist--fold dynamo
\citep{VZ72,CG95}, where a little bit of diffusion is needed to ``glue''
the constructively folded structures together and prevent this flux rope
arrangement from undoing itself.
The need for having magnetic diffusion in a dynamo
with an exponentially growing magnetic field in a steady
velocity field was also shown analytically in \cite{MP85}.
Without magnetic diffusion, on the other hand, there is the possibility
of solutions with progressively smaller characteristic length scales as
time goes on, but those cannot emerge from an eigenvalue problem.
In fact, an ideal magnetic field evolution with strictly vanishing
magnetic diffusivity can always be described in terms of the advection
of two Euler potentials, but no dynamo solutions with numerically resolved
Euler potentials have ever been found by this method \citep{Bra10}.
Nevertheless, the question regarding the need for finite magnetic
diffusion remains debated, as was demonstrated by a discussion during the
recent Nordita program on Solar Helicities in Theory and
Observations\footnote{\url{https://www.nordita.org/helicity2019}}.
It should be emphasized, however, that the problem of strictly vanishing
magnetic diffusivity is somewhat academic and disconnected with the
{\em limit} of vanishing magnetic diffusivity, in which case the existence
of what are known as fast dynamos has been proven \citep{Sow93,Sow94}.
In view of these complications, is it then still possible
to solve the dynamo problem with an ideal MHD code?
And even if it is possible, will the solution be wrong and if so,
in what way?

There is a related question about the use of ideal MHD in solving the dynamo problem.
Magnetic helicity is known to play an important role in certain types
of dynamos, namely those that amplify a large-scale magnetic field via
the $\alpha$ effect.  Such dynamos are driven by kinetic helicity.
This can produce a helical magnetic field, but since the magnetic
helicity is conserved by the ideal MHD equations, this happens in such
a way that there is magnetic helicity of opposite signs at different
length scales \citep{See96,Ji99}.
The question is therefore whether ideal MHD codes can describe this
evolution of magnetic helicity correctly.

Magnetic helicity conservation is an alien concept in numerical schemes
designed to solve the ideal MHD equations.
Such codes are primarily concerned with the conservation of mass,
momentum, energy, and magnetic flux.
Magnetic helicity, the volume integral of the magnetic field dotted
into its inverse curl, i.e., the magnetic vector potential, is not
normally considered.
At large magnetic Reynolds numbers or at high conductivity, magnetic
helicity changes only through fluxes \citep{BF84}.
Those can occur under inhomogeneous conditions or in the presence of
suitable boundary conditions.

Most code benchmarks are concerned with one- and two-dimensional
test problems. In those cases, the magnetic helicity vanishes from the outset.
We therefore need to resort to more complex three-dimensional problems
to see the effects of magnetic helicity and its dissipation properties.
A suitable benchmark that satisfies the aforementioned constraints
is the homogeneous helical dynamo problem in a periodic domain.
It produces large-scale magnetic fields through the $\alpha$ effect,
but the resulting magnetic helicity at large scales must have the
opposite sign to that of the kinetic helicity.
However, when the magnetic field at the wavenumber of the energy-carrying
eddies, $\kf$, reaches equipartition and saturates, the energy of the
large-scale magnetic field is still weak compared to the field at $\kf$.
The only way the large-scale magnetic field can grow further is by
dissipating magnetic helicity \citep{BF00}.
This should allow us to infer the rate of magnetic helicity dissipation.
The amplitude of the large-scale magnetic field is also controlled by
the evolution and destruction of magnetic helicity \citep{Bra01}.
This allows us to infer the effective scale dependence of the
numerical diffusion operator.

When magnetic helicity dissipation is accomplished through microphysical
resistivity, the dissipation rate is proportional to the current helicity.
The evolution of magnetic helicity is then given by
\begin{equation}
{\dd\over\dd t}\bra{\AAA\cdot\BB}=-2\eta\bra{\JJ\cdot\BB},
\label{dABdt}
\end{equation}
where $\BB=\nab\times\AAA$ is the magnetic field in terms of the magnetic
vector potential $\AAA$, $\JJ=\nab\times\BB$ is proportional to the
current density, and angle brackets denote volume averaging over a
periodic volume.
As can be seen from \Eq{dABdt}, the current helicity $\bra{\JJ\cdot\BB}$
must vanish on average once a statistically steady state is reached
\citep{Bra01}.
Again, this steady state is accompanied by a balance 
of large-scale and small-scale contributions of opposite signs.
Under isotropic conditions, the current helicity at a certain wavenumber
$k$ is equal to the spectral magnetic helicity times $k^2$, because the
former contains two more derivatives than the latter.
However, if magnetic helicity dissipation is accomplished through other
numerical processes, for example through hyperdiffusion, which has a
steeper dependence on the wavenumber, then this can affect the magnetic
helicity balance and therefore the final saturation value.
This was demonstrated numerically by \cite{BS02}.
Thus, a helically driven dynamo may be an excellent system to study
the properties of magnetic helicity dissipation, especially when this
is accomplished only through numerical processes.

It is useful to begin with models whose numerical resolution is relatively
small.  In fact, even a resolution of just $32^3$ mesh points is enough to find
large-scale dynamo action; see \cite{Bra01} for early models of that type.
His simulations showed that at larger magnetic Reynolds numbers,
and thus at higher resolution, it takes progressively longer to reach the
final saturation state of such a system with periodic boundary conditions.
Simulations at higher resolution therefore increase the risk of not
noticing that a solution has not yet reached its final state.

In the present paper, we first motivate and describe the details of our
model (Section 2), and then present the results for the magnetic field
evolutions at different numerical resolutions (Section 3) and compare
in some cases with results of direct numerical simulations (DNS; see
Sections 3.4 and 3.5).
We present our discussion in Section~4 and finish with concluding remarks
in Section~5.

\section{The model}

\subsection{Periodic boundary conditions}

We consider here the arguably simplest setup of a large-scale turbulent
dynamo.
We drive turbulence through helical isotropic random forcing, which
leads to an $\alpha$ effect.
It is responsible for driving what in a sphere would be called poloidal
and toroidal fields.
Because the $\alpha$ effect acts in both steps, the resulting system is
called an $\alpha^2$ dynamo.
We adopt periodic boundary conditions, as is commonly done in numerical
studies of hydrodynamic and MHD turbulence.

We should emphasize from the outset that it is this assumption of
periodicity that is primarily responsible for causing features of this
dynamo that would not occur in astrophysical setups, namely the
generation of a superequipartition magnetic field and a resistively
slow evolution toward this final state \citep{Bra01}.
In real systems that are not periodic, magnetic helicity fluxes are
believed to be important in high magnetic Reynolds number turbulence
\citep{BF00}.
Those fluxes can prevent a resistively slow evolution while still allowing
the system to saturate at approximately the equipartition field strength
\citep{Bra18}.
Here, however, we are interested in quantifying the extent to which
non-ideal effects play a role in an ideal MHD code, and so periodic boundary conditions
are appropriate.

\subsection{Setup of the model}

We adopt a cubic domain of side length $L=1$,
so the smallest wavenumber in the domain is $k_1=2\pi$.
We solve the compressible MHD equations with a forcing function $\ff$
on the right-hand side of the momentum equation.
This forcing function is random in time and has a
characteristic wavenumber $\kf$ that we choose to be larger than
$k_1$ by a certain factor.
The forcing function has positive helicity, so
$\bra{\ff\cdot\nab\times\ff}/\kf\bra{\ff^2}$ is positive
and close to unity.

\subsection{Code and choice of parameters}

We use \FLASHz\footnote{\url{http://flash.uchicago.edu/site/flashcode/}}
\citep{FLASH}, to solve the equations for an isothermal gas, choosing an ideal
gas with a $\gamma=1$ equation of state.
The sound speed is unity, so the  root mean square (rms) value of the
velocity $\uu$ is automatically equal to the Mach number.
We force the flow such that it remains subsonic on average with
$\urms\approx0.3$.

We use Lorentz--Heaviside units, so the mean magnetic energy density is given by
$\EEM=\bra{\BB^2}/2$.
The density $\rho$ is initially unity.
Furthermore, because no mass enters or leaves the domain,
the mean density remains always unity.

We use the MHD eight-wave module of \FLASH \citep{Derigs16},
which is based on a divergence-cleaning algorithm.
The forcing function is analogous to that used by \cite{SPS14},
except that here only one sign of helicity is used.
In particular, we used an artificial forcing term $F$ which is modeled
as a stochastic Ornstein--Uhlenbeck process \citep{Eswaran88,Benzi08} 
with a user-specified forcing correlation time, which was taken to be one half.  
In the following, we consider two values for the scale separation
ratio $\kf/k_1$: a smaller one with a combination of 76 wavevectors
with wavenumbers between 2 and 3, and a larger one with 156 wavevectors
with wavenumbers between 4 and 5.
These cases are distinguished by their average nominal forcing wavenumbers
of 2.5 and 4.5, respectively.
Relevant additions to \FLASH that were used for the present studies
are being provided through the online material \citep{BS19}, which also
contains input data and analysis tools used for the figures of this paper.

\section{Results}

\subsection{Weak scale separation}

\begin{figure}[t!]\begin{center}
\includegraphics[width=\columnwidth]{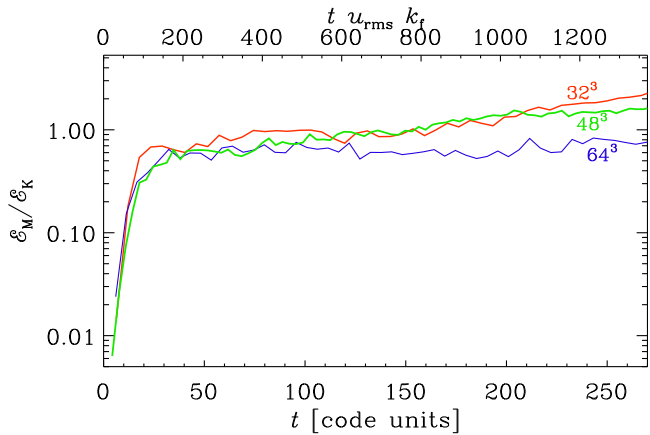}
\end{center}\caption[]{
Early evolution of the normalized magnetic energy
for resolutions $32^3$, $48^3$, $64^3$ and $\kf=2.5$.
The upper abscissa gives time in eddy turnover times based on the run with $48^3.$
}\label{pcomp_early}\end{figure}

In \Fig{pcomp_early}, we plot the growth of $\EEM$, normalized by
the long-time average (indicated by a subscript $t$, evaluated
during the saturated phase of the dynamo) of the mean kinetic energy density,
$\EEK=\bra{\rho\uu^2}_t/2$, for different numerical resolutions.
Time is given both in code units and in eddy turnover times, $(\urms\kf)^{-1}$.
Ignoring density fluctuations, we define $\urms=(2\EEK)^{1/2}$.
In all cases, the initial exponential growth phase is the same and the
growth rate of the rms magnetic field (proportional to $\EEM^{1/2}$)
is $\lambda\approx0.18$ in code units, corresponding to
$\lambda/\urms\kf\approx0.036$ in units of the turnover rate.
The magnetic energy saturates approximately at the equipartition level
with $\EEM\approx\EEK$.

\begin{figure}[t!]\begin{center}
\includegraphics[width=\columnwidth]{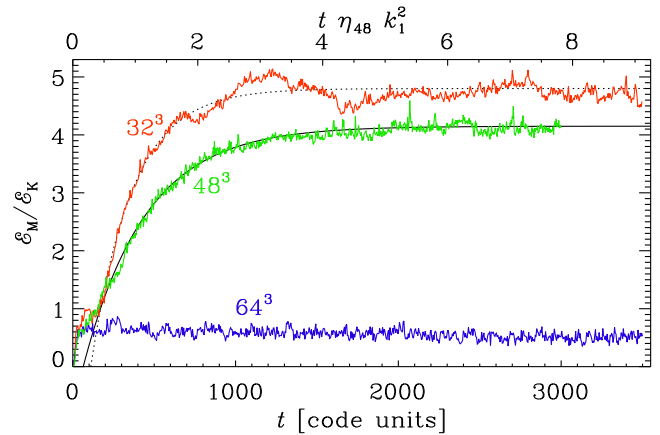}
\end{center}\caption[]{
Saturation for resolutions $32^3$, $48^3$, $64^3$ and $\kf=2.5$.
The upper abscissa gives time in microphysical diffusion times
based on the run with $48^3$.
The dotted line gives the fit, as explained in the text.
}\label{pcomp}\end{figure}

The magnetic field evolution shown in \Fig{pcomp_early} covers only the
early saturation phase.
At later times, the magnetic energy continues to increase for two of
the runs, as shown in \Fig{pcomp}.
In fact, the system reaches values that exceed $\EEK$ by a factor of 4--5.

Following \cite{Bra01}, we fit the late-time evolution of the magnetic
energy to a curve of the form
\begin{equation}
\EEM-\EEK\approx\EEK\,{\kfeff\over k_1}
\left[1-e^{-2\eta k_1^2(t-t_{\rm sat})}\right]
\quad\mbox{for $t>t_{\rm sat}$,}
\label{EEMsat}
\end{equation}
where $\kfeff$ and $t_{\rm sat}$ are fit parameters that characterize
the effective forcing wavenumber and the effective saturation time,
respectively (see \App{Late} for a derivation).
In the simulations in which $\eta$ is formally zero, we also
replace $\eta$ by $\eta^{\rm eff}$ as an effective parameter that can
be obtained from a fit to the evolution of $\EEM(t)$.
These parameters are listed in \Tab{Tsum}, along with other parameters
characterizing the simulations.
In particular, we also compare with the estimated {\em turbulent}
magnetic diffusivity, $\etatz=\urms/3\kf$ \citep[see, e.g.][]{BB02}.
The ratio $3\etatz/\eta^{\rm eff}$ corresponds to the magnetic Reynolds
number.
In a few cases, however, we also add an explicit magnetic diffusivity;
see the column denoted in \Tab{Tsum} by $\eta_{-6}^{}$.
Those runs will be discussed separately in \Sec{WithExplicit}.

\begin{table}[b!]\caption{
Parameters of the various runs.
}\vspace{12pt}\centerline{\begin{tabular}{lccccccccc}
Res & $\kf/k_1$ & $\urms$ & $\kfeff$ & $t_{\rm sat}$ & 
$\eta_{-6}^{}$ & $\eta_{-6}^{\rm eff}$ &
$3\etatz/\eta^{\rm eff}$\\
\hline
$32^3$ & 2.5 & 0.28 & 3.8 & 170 &  0 & $50$ & 360 \\
$48^3$ & 2.5 & 0.30 & 3.2 & 170 &  0 & $66$ & 270 \\
$64^3$ & 2.5 & 0.30 & --- & --- &  0 &  --- & --- \\
$64^3$ & 2.5 & 0.25 &11.7 &   5 &  5 & $34$ & 470 \\
$32^3$ & 4.5 & 0.28 & 8.0 &  60 &  0 &$100$ & 150 \\
$48^3$ & 4.5 & 0.29 & --- & --- &  0 & ---  & --- \\
$64^3$ & 4.5 & 0.30 & --- & --- &  0 &  --- & --- \\
$64^3$ & 4.5 & 0.25 &12.1 &   5 &  5 &$ 64$ & 140 \\
$64^3$ & 4.5 & 0.21 & 4.7 &  10 & 50 &$150$ &  49 \\
\label{Tsum}\end{tabular}}
\tablenotemark{
All quantities are in code units; $\eta_{-6}^{}$ and
$\eta_{-6}^{\rm eff}$ denote values in units of $10^{-6}$.
}\end{table}

As we see from \Tab{Tsum}, the value of $\kfeff$ does not vastly exceed
the nominal value of $\kf$.
This is somewhat surprising, given that one would have expected
that the numerical diffusion operator might be more efficient at
high wavenumbers, as is the case with hyperdiffusion; see
the corresponding numerical experiments of \cite{BS02}.
This is apparently not the case.
In some of the runs with explicit diffusion, however, there are cases
where $\kfeff$ exceeds the nominal value of $\kf$ by a factor of 3--5.

There are two more fit parameters.
One is $\eta^{\rm eff}$, which is inferred from a fit to the
saturation behavior given by \Eq{EEMsat}.
Its values are found to be small by comparison with the product
$\urms\delta x\approx5\times10^{-3}$, where $\delta x=1/32$ is the
mesh spacing.
The other fit parameter is $t_{\rm sat}$, whose values are listed for
completeness; they characterize merely the time when the early saturation
phases ends and this depends also on the value of the initial field.
It is therefore not a parameter characterizing the numerical diffusion
scheme.
It turns out to be about the same for the $48^3$ and $32^3$ runs.

\begin{figure}[t!]\begin{center}
\includegraphics[width=.32\columnwidth]{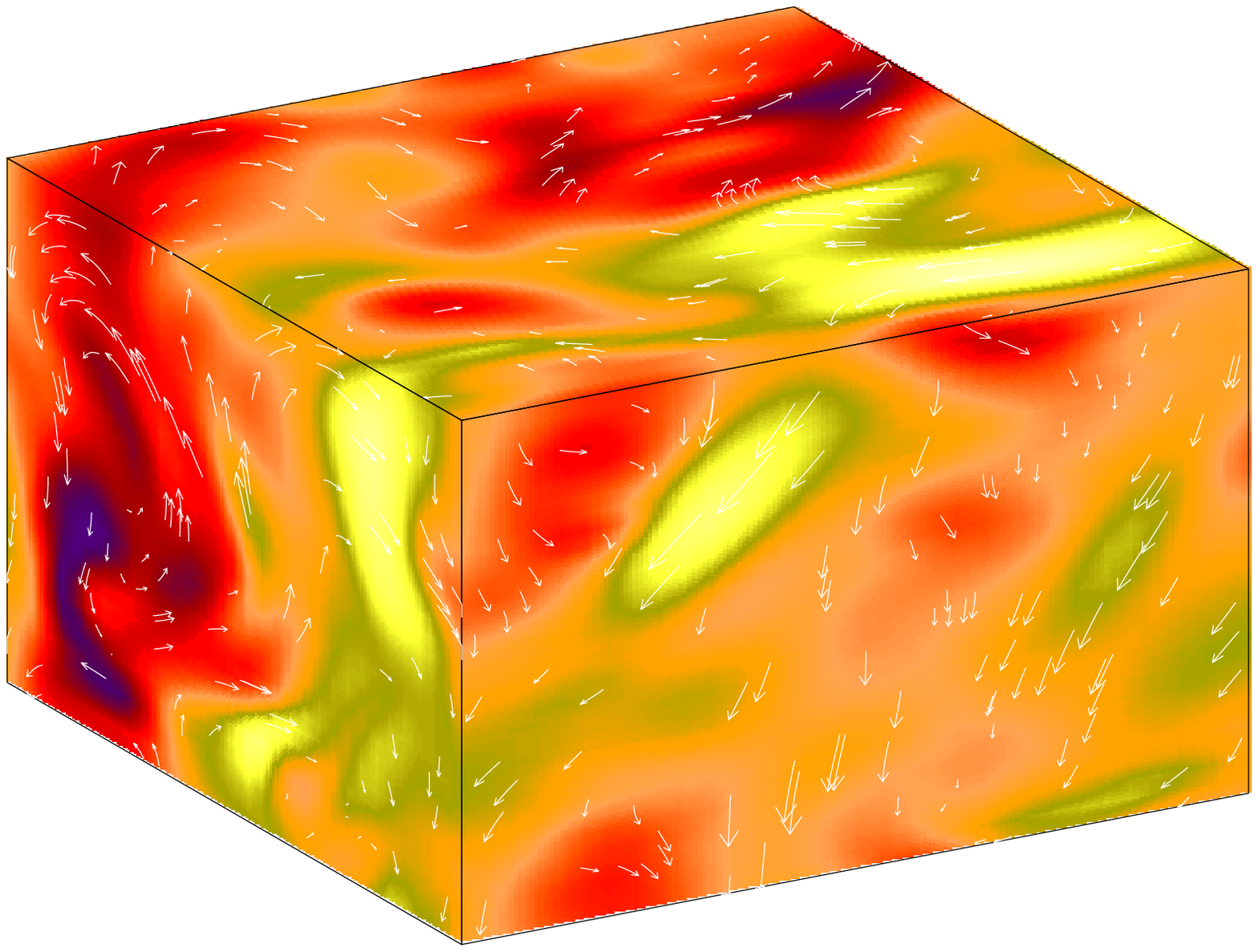}
\includegraphics[width=.32\columnwidth]{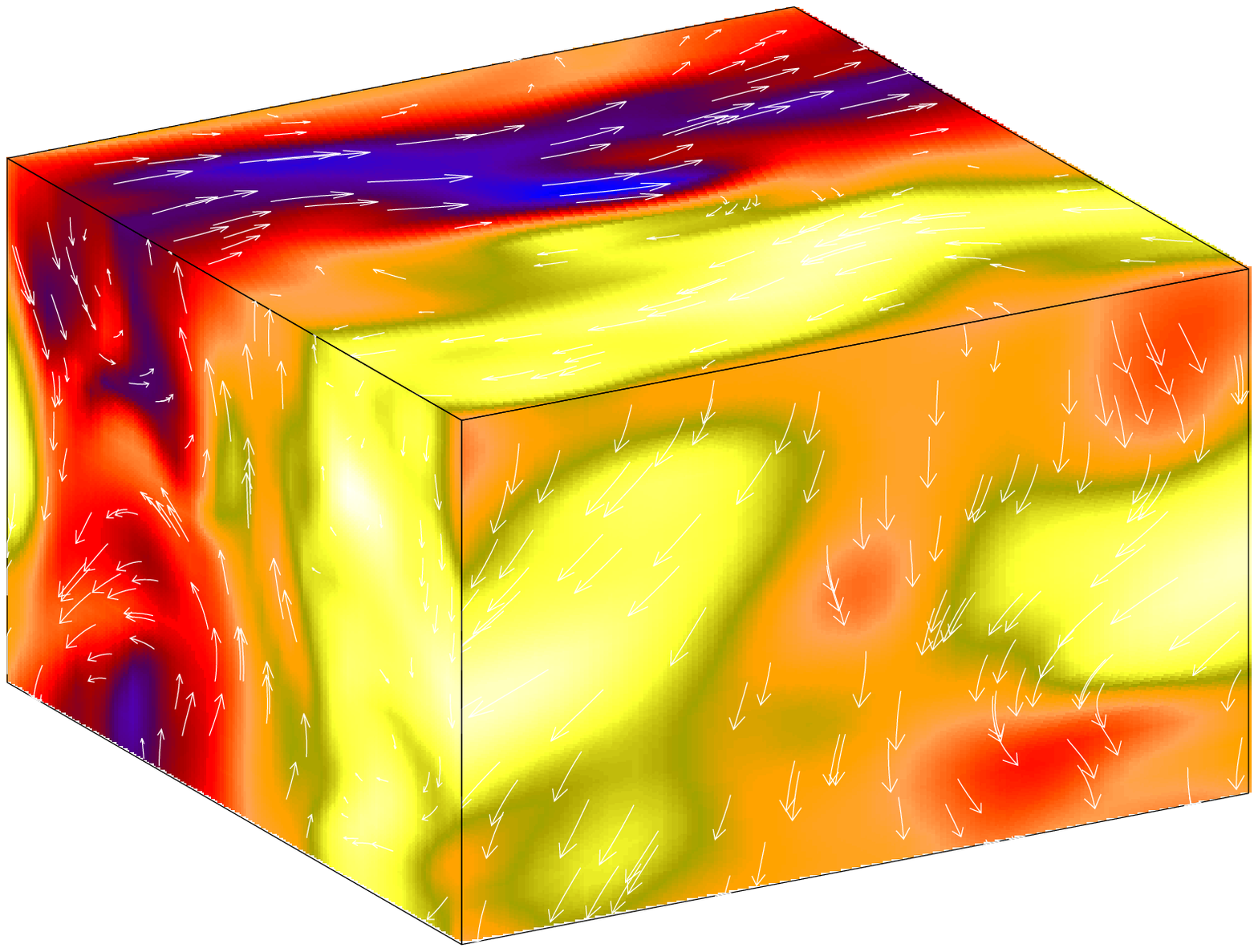}
\includegraphics[width=.32\columnwidth]{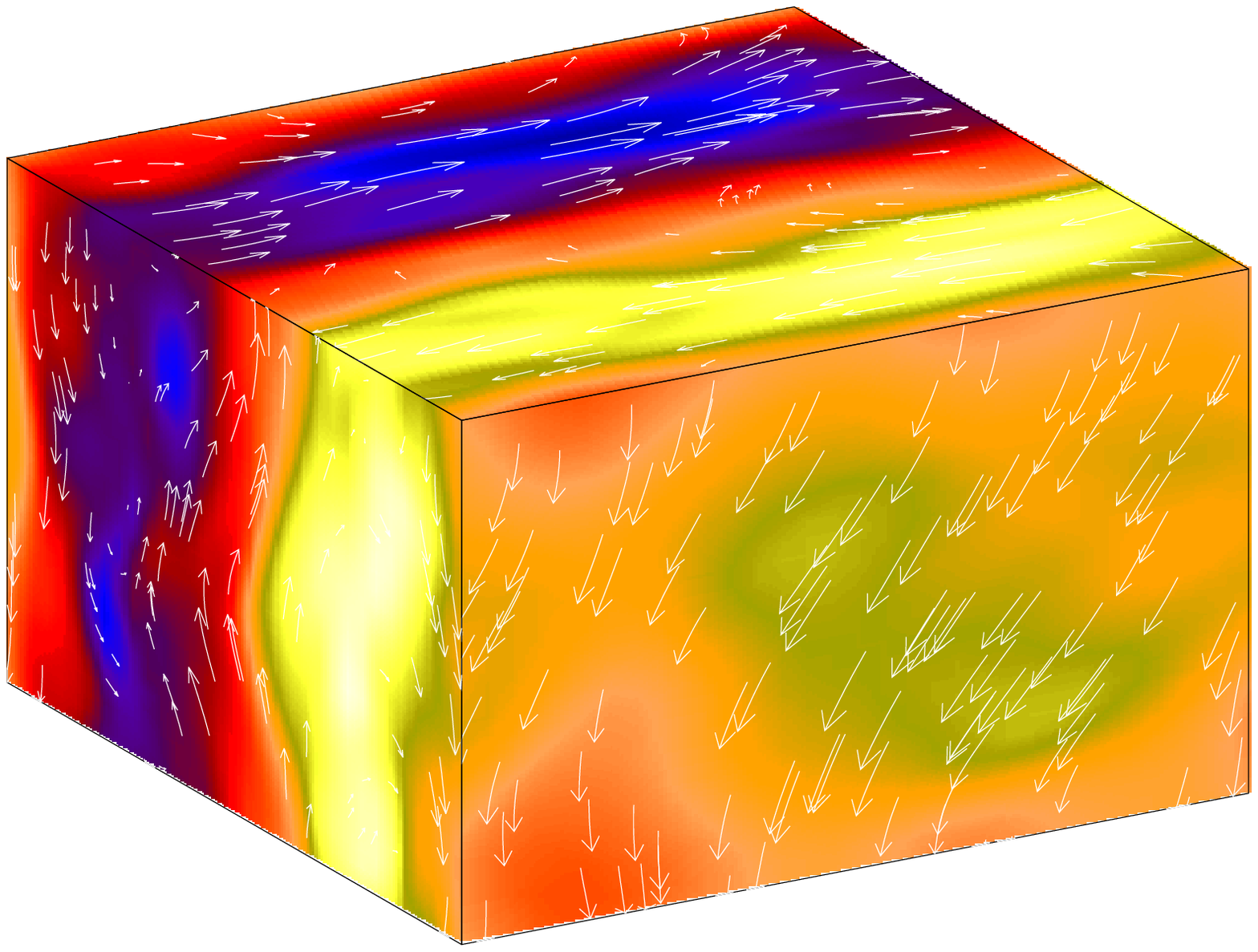}
\end{center}\caption[]{
Visualizations of $B_x$ and vectors of $\BB$ (in white) on the periphery
of the domain at times 200, 300, and 3500
for $\kf/k_1=2.5$ with $32^3$ mesh points.
Yellow (blue) shades denote positive (negative) values.
}\label{bx_Helix_hdf5_chk_0800}\end{figure}

In \Fig{bx_Helix_hdf5_chk_0800}, we show a visualization of
$B_x$ on the periphery of the computational domain at selected
times during the late saturation phase.
We see that, at late times, $B_x$ shows a sinusoidal variation in the
$y$ direction.
There is also a similar variation of $B_z$, but it is phase shifted by
$90\degr$ relative to $B_x$ and not shown here.
This type of field structure is one of three possible field
configurations that all have negative magnetic helicity; see
\cite{Bra01} for details.

\begin{figure}[t!]\begin{center}
\includegraphics[width=\columnwidth]{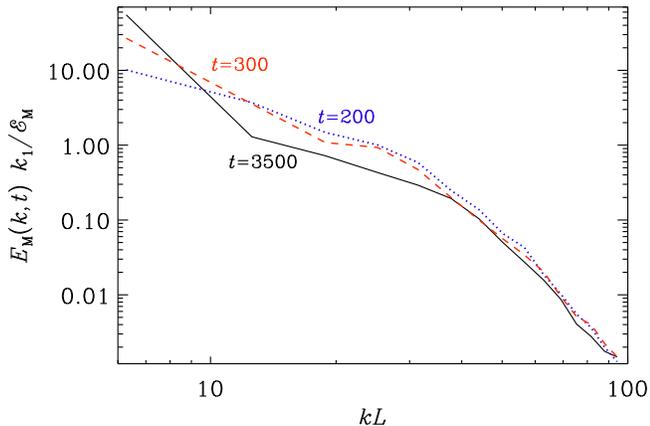}
\end{center}\caption[]{
Magnetic energy spectra at times 200, 300, and 3500
for $\kf/k_1=2.5$ with $32^3$ mesh points.
}\label{spec_Helix_hdf5_chk_0300}\end{figure}

\begin{figure}[t!]\begin{center}
\includegraphics[width=.96\columnwidth]{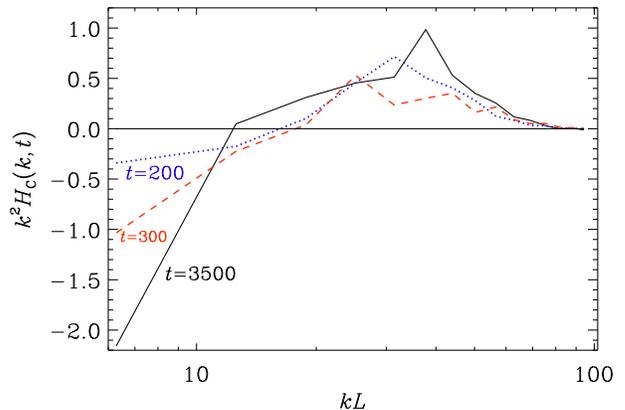}
\end{center}\caption[]{
Scaled current helicity dissipation spectra, $k^2\HC(k,t)$, at times 200, 300,
and 3500 for $\kf/k_1=2.5$ with $32^3$ mesh points.
}\label{pspecJBcomp}\end{figure}

In \Fig{spec_Helix_hdf5_chk_0300}, we show magnetic energy spectra,
$\EM(k,t)$, at different times.
They are normalized such that
\begin{equation}
\int_0^\infty\EM(k,t)\,\dd k=\EEM(t),
\end{equation}
is the mean magnetic energy density.
We clearly see that most of the magnetic energy is at the smallest
possible wavenumber, $k=k_1$, corresponding to the largest possible
scale of the system.
In this case, the spectra show no particular feature at the forcing
wavenumber.
This may partly be caused by the relatively poor scale separation
ratio, i.e., $\kf$ is not very large compared to $k_1$.
Another reason may be the small resolution of only $32^3$ mesh points.
The largest wavenumber in the domain is the Nyquist wavenumber,
$k_{\rm Ny}=\pi/\delta x=\pi N/L\approx50$ for this resolution
with $N=32$ mesh points per direction,
and $\approx100$ for $N=64$ mesh points.
Corresponding current helicity spectra, $\HC(k,t)$, scaled with $k^2$,
are shown in \Fig{pspecJBcomp}.
Note that $\HC(k,t)$ is normalized such that
$\int\HC\,\dd k=\bra{\JJ\cdot\BB}$, where $\JJ=\nab\times\BB$ is
proportional to the current density.
The scaling with $k^2$ has been adopted so that the high wavenumber part
of the spectrum can be seen more clearly.
Theoretically, however, we would have expected that, at late times,
$\bra{\JJ\cdot\BB}=0$, so that the positive and negative parts of
$\HC$ should cancel, but not those of $k^2\HC$; see \App{Late}.

\begin{figure}[t!]\begin{center}
\includegraphics[width=.96\columnwidth]{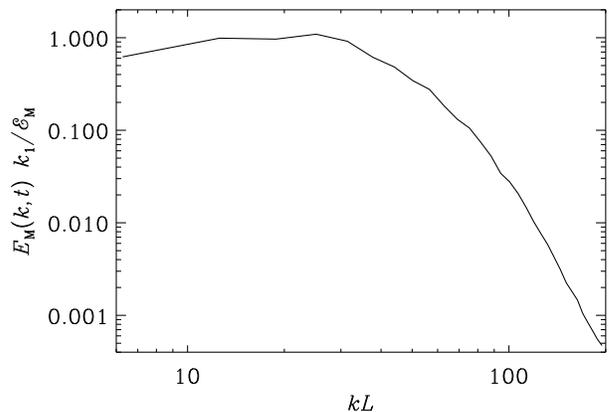}
\end{center}\caption[]{
Spectrum for the higher resolution run with $64^3$ mesh points
at time 3500, i.e., the end of the run,
for $\kf/k_1=2.5$.
}\label{spec_Helix_hdf5_chk_0440}\end{figure}

Our higher resolution run with $64^3$ mesh points does not develop
a large-scale magnetic field.
The resulting magnetic energy spectrum is shown in
\Fig{spec_Helix_hdf5_chk_0440}.
The magnetic energy spectrum is seen to peak at $kL\approx30$, which
corresponds to $k/k_1\approx5$.
This is twice as large as the value of $\kf/k_1=2.5$.
Such behavior is typical of small-scale dynamo action \citep{Scheko04}.

\subsection{Larger scale separation ratio}

\begin{figure}[t!]\begin{center}
\includegraphics[width=\columnwidth]{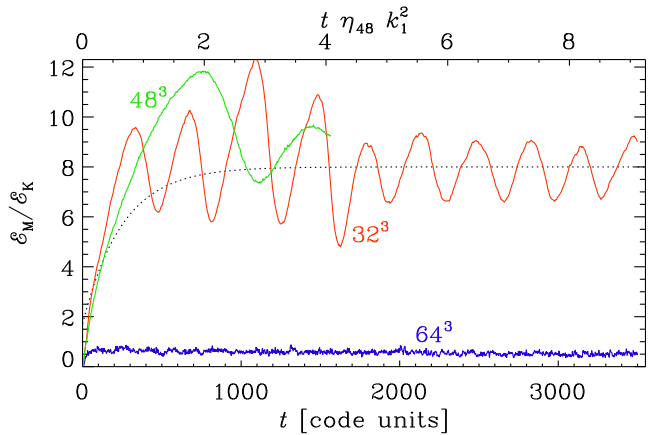}
\end{center}\caption[]{
Late saturation for resolutions $32^3$, $48^3$, $64^3$ and $\kf/k_1=4.5$.
The upper abscissa gives time in microphysical diffusion times based on
the empirical value $\eta_{48}$ found for the run with $48^3$ mesh points.
The dotted line gives an attempted fit.
}\label{pcomp_highk}\end{figure}

We have increased the value of $\kf$ to include wavenumbers between
4 and 5.
This scale separation ratio is still not very large, but we should
keep in mind that the resolution is not very large either, and
$k_{\rm Ny}/k_1$ is only 16 for our $32^3$ simulations.
The results turn out to be quite different in many ways:
first, the mean magnetic energy density shows oscillatory behavior
(\Fig{pcomp_highk}) and second, the magnetic field develops a large-scale
component already very early on.
This behavior is rather unexpected.
We also see that in the kinematic phase, the magnetic energy grows
slightly faster than in the case of a smaller scale separation ratio.
For the run with $64^3$ mesh points, there is again no large-scale dynamo.
Furthermore, normalized by the kinetic energy, the magnetic energy
generated by the small-scale dynamo is now about half as strong as
in the case with $\kf/k_1=2.5$.
This can be explained by the fact that the effective magnetic Reynolds
number based on the value of $\kf$ is now smaller.

\begin{figure}[t!]\begin{center}
\includegraphics[width=.98\columnwidth]{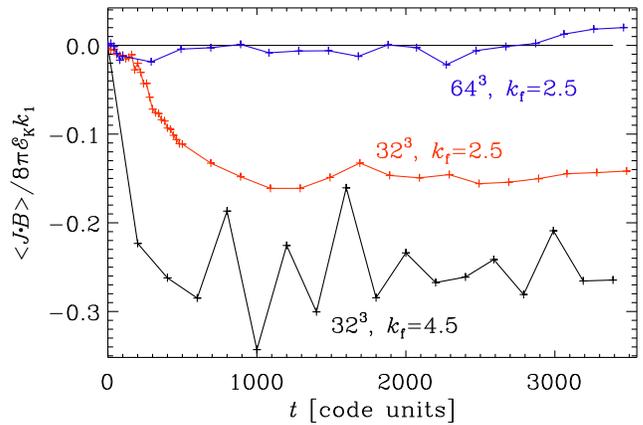}
\end{center}\caption[]{
Evolution of current helicity for runs with different resolutions
($32^3$, $64^3$) and different scale separation ($\kf/k_1=2.5$ and $4.5$).
}\label{pjbm}\end{figure}

In \Fig{pjbm}, we show the evolution of current helicity,
$\bra{\JJ\cdot\BB}$ for runs with different resolutions
($32^3$, $64^3$) and different scale separation ($\kf/k_1=2.5$ and $4.5$).
Except for the run with $64^3$ mesh points and $\kf/k_1=2.5$, where
$\bra{\JJ\cdot\BB}$ is seen to fluctuate around zero, we find a clear
evolution away from zero with subsequent saturation at a negative value
for the other two runs.
It is therefore clear that the numerical evolution of magnetic helicity is
-- unlike the proper resistive case -- not simply controlled by the value
of the current helicity, because a finite value of $\bra{\JJ\cdot\BB}$
should continue to drive magnetic helicity, $\bra{\AAA\cdot\BB}$, to a
new state all the time; see \App{Late}.

\begin{figure}[t!]\begin{center}
\includegraphics[width=.98\columnwidth]{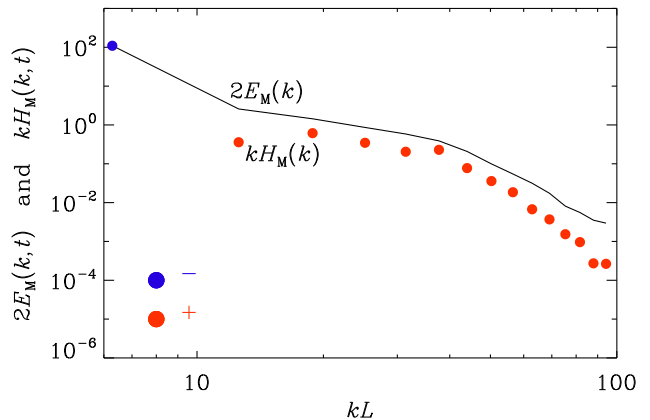}
\end{center}\caption[]{
Comparison of magnetic energy and scaled helicity spectra for the run with
$32^3$ mesh points and $\kf/k_1=2.5$ at $t=3500$.
Positive (negative) values of $\HM$ are plotted as red (blue) symbols.
}\label{pspec}\end{figure}

To compute magnetic helicity spectra, $\HM(k,t)$, we make use of the
fact that, under homogeneous conditions, $\HM(k,t)$ is related to the
current helicity spectrum $\HC(k,t)$ via $\HM(k,t)=\HC(k,t)/k^2$.
For the spectrum shown in \Fig{pspec}, we have verified this relation
by computing $\HM(k,t)$ directly from $\AAA$ in Fourier space (indicated
by tildes) as $\tildeA_i=\epsilon_{ijl}\,\ii k_j\tildeB_l/k^2$ in the Coulomb gauge.
It is normalized analogously to $\HC$ as
$\int\HM(k,t)\,\dd k=\bra{\AAA\cdot\BB}$.
In \Fig{pspec}, we compare the magnetic energy with the scaled magnetic
helicity spectrum for the run with $32^3$ mesh points and $\kf/k_1=2.5$
at $t=3500$ (in code units).
We see that the spectral magnetic helicity is negative for $k=k_1$ and
positive for all larger values of $k$.

\begin{figure}[t!]\begin{center}
\includegraphics[width=.48\columnwidth]{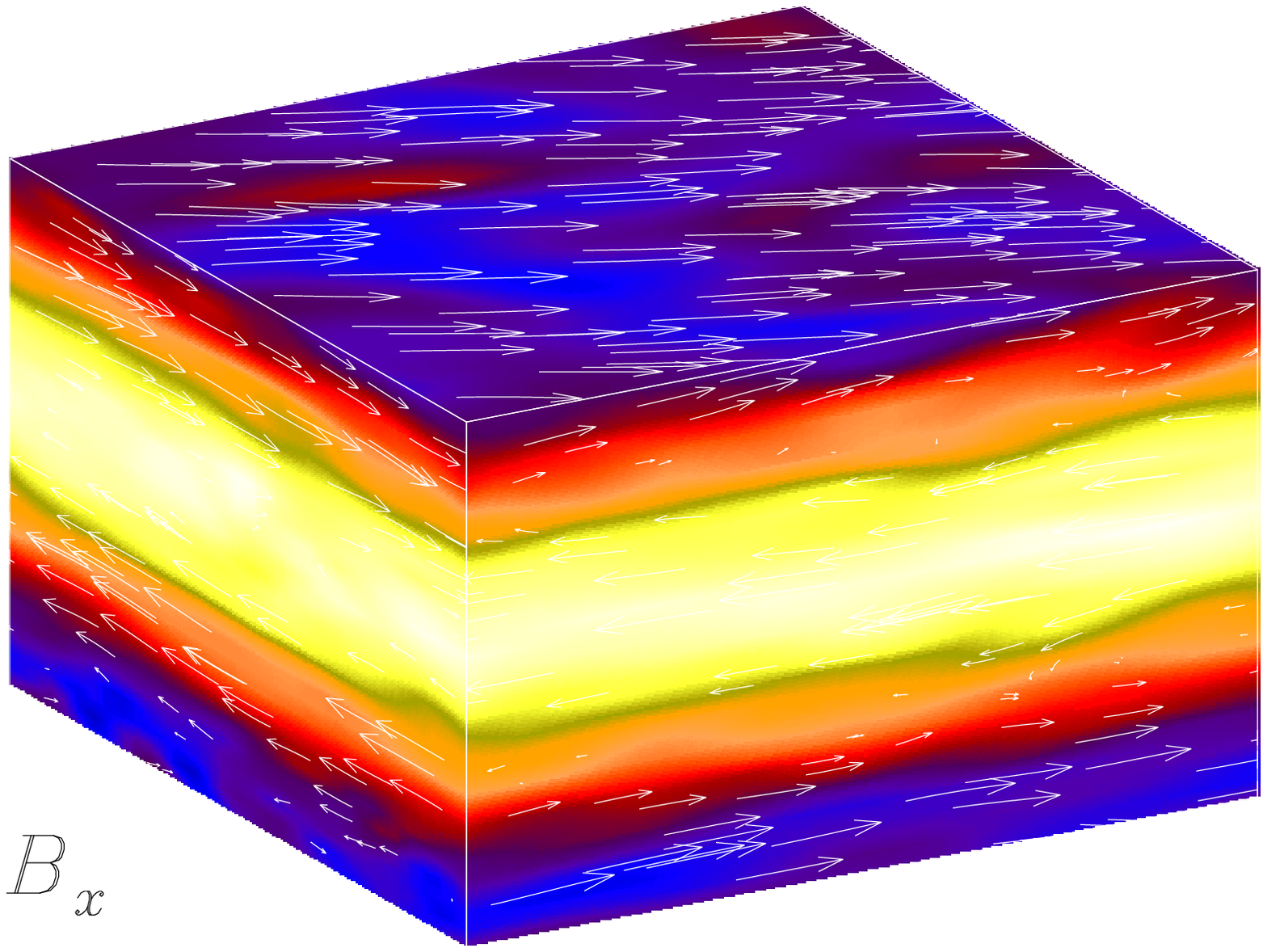}
\includegraphics[width=.48\columnwidth]{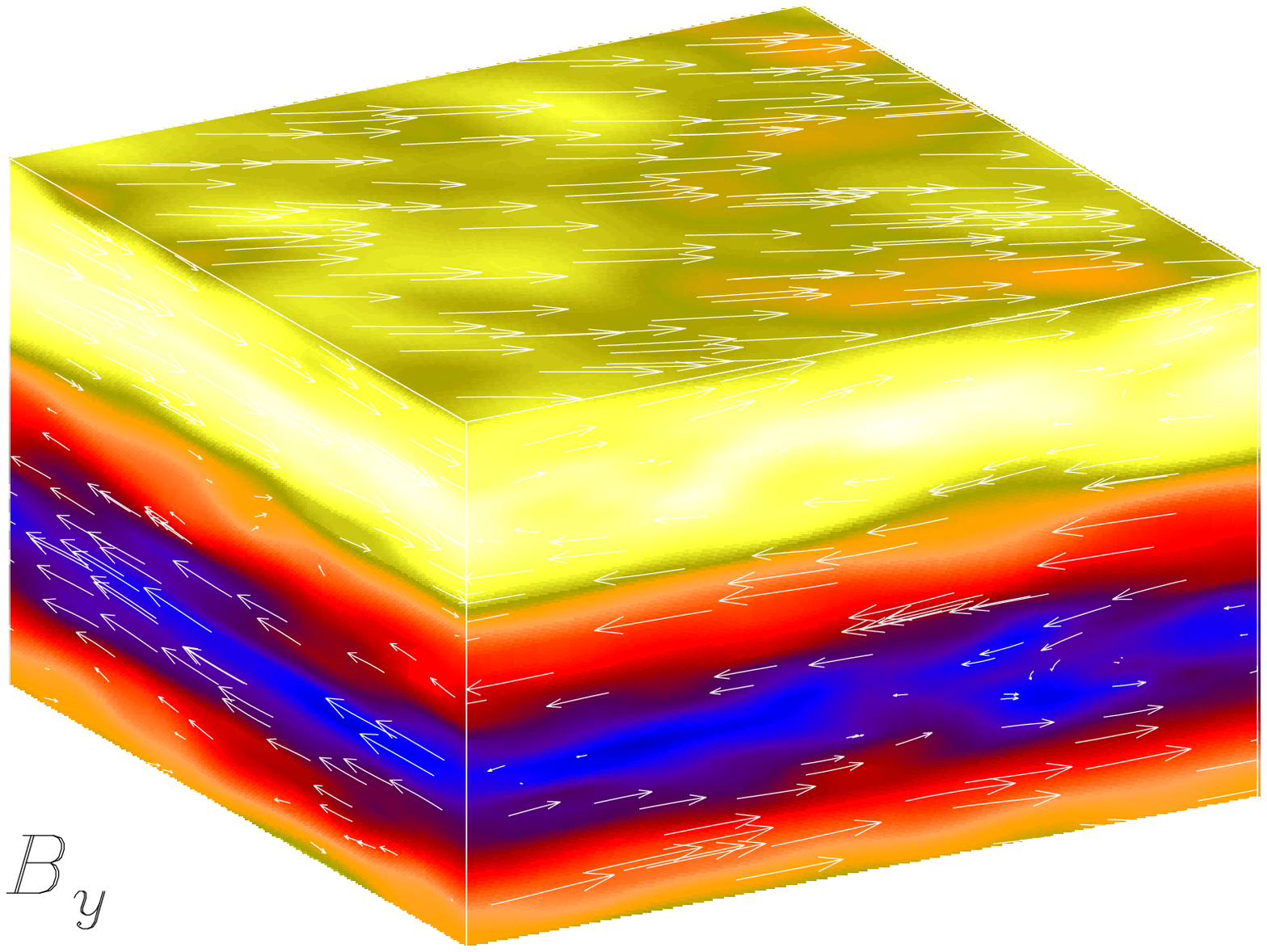}
\end{center}\caption[]{
$B_x$ and $B_y$ at time 200 (in code units) for $\kf/k_1=4.5$
with $32^3$ mesh points.
Note that the fields now vary with $z$, and that the phases of
$B_x$ and $B_y$ are shifted by $90\degr$ relative to each other.
Yellow (blue) shades denote positive (negative) values.
}\label{bx_Helix_hdf5_chk_0340}\end{figure}

In \Fig{bx_Helix_hdf5_chk_0340}, we show visualizations of $B_x$ and $B_y$
for $\kf/k_1=4.5$ and $32^3$ mesh points.
A large-scale magnetic field develops very quickly.
Unlike the case shown in \Fig{bx_Helix_hdf5_chk_0800}, the mean
magnetic field now varies in the $z$ direction and is here,
except for an insignificant overall phase shift, of the form
$\meanBB=(\cos k_1z, \sin k_1z, 0)$.

\subsection{Runs with explicit magnetic diffusivity}
\label{WithExplicit}

\begin{figure}[t!]\begin{center}
\includegraphics[width=\columnwidth]{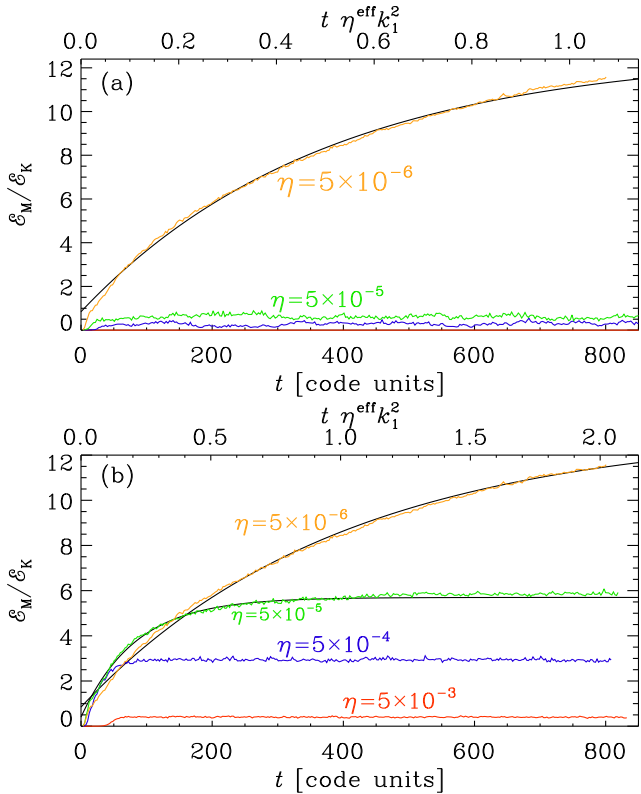}
\end{center}\caption[]{
Saturation for runs with explicit magnetic diffusivity
using (a) $\kf/k_1=2.5$ and (b) $\kf/k_1=4.5$ with
$\eta=5\times10^{-3}$ (red), $5\times10^{-4}$ (blue),
$5\times10^{-5}$ (green), and $5\times10^{-5}$ (orange), 
all at a resolution of $64^3$ mesh points.
The upper abscissa gives time in effective microphysical diffusion
times based on the runs with the largest saturation value.
The black solid lines represent the fits.
}\label{pcomp64_highk_eta5}\end{figure}

\FLASH allows for the possibility of adding an explicit magnetic
diffusivity $\eta$.
We now present simulations using for $\eta$ the same value as the effective one of
$5\times10^{-5}$ found in the $32^3$ simulations with $\kf/k_1=2.5$.
In this case we carry out simulations with $64^3$ mesh points, where previously no
large-scale magnetic field was found with \FLASHz.
We also include a run with $\eta=5\times10^{-6}$.
In \Fig{pcomp64_highk_eta5}, we show the results for $\kf/k_1=2.5$
and $4.5$.

It turns out that there is large-scale magnetic field growth in the
case with $\kf/k_1=4.5$ and $\eta=5\times10^{-4}$ or less, but not for
$5\times10^{-3}$ or more, and also not for $\kf/k_1=2.5$.
In both cases, however, there is large-scale dynamo action with
$\eta=5\times10^{-6}$.
Interestingly, the value of $\eta^{\rm eff}$ is always larger than that
of $\eta$ by a factor of 3--13; see \Tab{Tsum}.

To understand the absence of large-scale dynamo action for $\kf/k_1=2.5$
and $\eta=5\times10^{-5}$, we must remember that $\kf/k_1$ must exceed
a certain limit, which \cite{HBD04} found to be around 2.2; see their
Figure~23.
Whether the smallness of $\kf$ is indeed the reason for the absence of
dynamo action in our case with $\kf/k_1=2.5$ cannot be conclusively answered
and requires more dedicated tests with the {\sc Pencil Code}, which are described next.

\subsection{Comparison with the {\sc Pencil Code}}

We now compare with DNS results obtained with the {\sc Pencil
Code}.\footnote{\url{https://github.com/pencil-code}}.
Again, we use $\eta=5\times10^{-5}$ along with our two values of
$\kf/k_1$, namely 2.5 and 4.5.
In both cases, we find large-scale dynamo action.
As expected, the amplitudes are different; compare the values of
$\kfeff$ for the different values of $\kf$ in \Tab{Tsum2}.
The kinematic growth rate varies between $\lambda=0.15$ and $0.30$,
which is compatible with the value of $0.18$ obtained with \FLASHz.

\begin{table}[b!]\caption{
Parameters of runs with the {\sc Pencil Code}.
}\vspace{12pt}\centerline{\begin{tabular}{lccccccccc}
Res & $\kf/k_1$ & $\Pm$ & $\urms$ & $\kfeff$ & $t_{\rm sat}$ & 
$3\etatz/\eta$\\
\hline
$32^3$ & 2.2 & 10 & 0.11 & 1.40& 120 & 81 \\
$32^3$ & 2.6 & 10 & 0.11 & 1.76& 110 & 81 \\
$32^3$ & 4.5 & 10 & 0.11 & 3.20&  70 & 78 \\
$64^3$ & 4.5 & 10 & 0.12 & 3.86&  90 & 82 \\
$64^3$ & 4.5 & 20 & 0.10 & 4.15& 105 & 70 \\
$64^3$ & 4.5 & 40 & 0.08 & 4.20& 150 & 55 \\
\label{Tsum2}\end{tabular}}
\tablenotemark{
In all cases, $\eta^{\rm eff}=\eta=5\times10^{-5}$,
and $\Rm=3\etatz/\eta$.
}\end{table}

\begin{figure}[t!]\begin{center}
\includegraphics[width=\columnwidth]{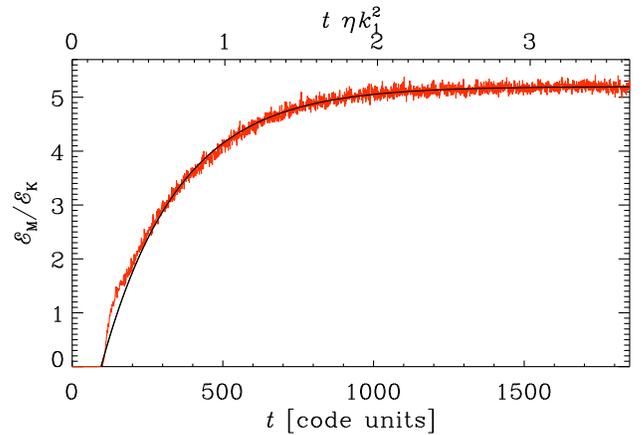}
\end{center}\caption[]{
Direct numerical simulations with the {\sc Pencil Code} using
$\eta=5\times10^{-5}$ and $\kf/k_1=4.5$.
}\label{psat_pencil}\end{figure}

Given that we perform DNS without subgrid scale modeling, there is a
limit to the smallest value of the viscosity $\nu$ that can be used at
the resolutions adopted here, which are $32^3$ or $64^3$ mesh points.
It turns out that in all cases with $\eta=5\times10^{-5}$ and
$\nu=5\times10^{-4}$, the code produces acceptable results for $t\la2000$
time units, but the code crashes at later times.
This problem disappears when the viscosity is increased to
$\nu=2\times10^{-3}$, while $\eta=5\times10^{-5}$ is kept unchanged.
The evolution of $\EEM/\EEK$, as obtained with the {\sc Pencil Code},
is shown in \Fig{psat_pencil}.
The corresponding values of the magnetic Prandtl number,
$\Pm\equiv\nu/\eta$, are given in \Tab{Tsum2}.
We see that the results for $\kfeff$ are not very sensitive to the
value of $\nu$.

It is important to realize that in DNS, there is no separate $\eta^{\rm eff}$,
because the coefficient entering in \Eq{EEMsat} is always the same as
the input parameter $\eta$ used.
In all cases, the fit works well and there is no spurious diffusivity
entering the resistively slow saturation phase.
This is different in the \FLASH code, where $\eta^{\rm eff}$ tends to
exceed $\eta$ by a factor of 3--13.

\begin{figure}[t!]\begin{center}
\includegraphics[width=\columnwidth]{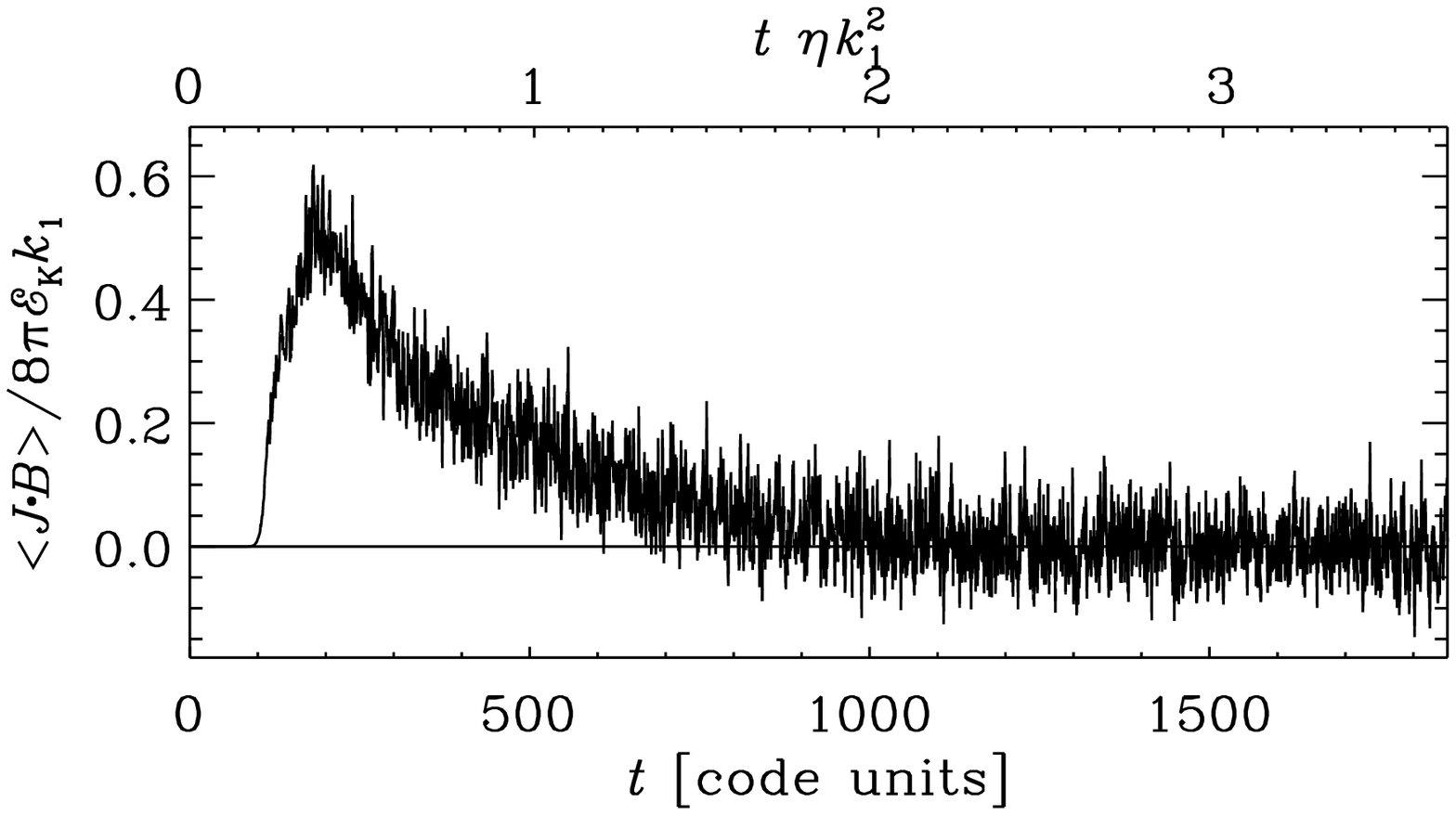}
\end{center}\caption[]{
Evolution of $\bra{\JJ\cdot\BB}$ for the run of \Fig{psat_pencil}
using the {\sc Pencil Code}.
}\label{pjbm_pencil}\end{figure}

As discussed above, $\bra{\JJ\cdot\BB}$ should approach zero at late
times.
This is shown in \Fig{pjbm_pencil}, which demonstrates that
$\bra{\JJ\cdot\BB}$ is initially zero, begins to rise after about 100
time units, reaches then a positive maximum after about one third of a
diffusion time, and then decays to zero on a resistive time scale.
It is interesting to note that $\bra{\JJ\cdot\BB}$ is positive,
while in the ideal simulations with \FLASHz, it has a negative value;
see \Fig{pjbm}.

Looking at the corresponding magnetic energy spectrum of
\Fig{pspec} with \FLASHz, we see that there is a strong dominance of
the large-scale field over the small-scale field.
This is also consistent with the corresponding current helicity spectra
shown in \Fig{pspecJBcomp}, keeping in mind that we scaled $\HC(k,t)$
with $k^2$ to show the rather weak contributions from small scales.
Thus, we can conclude that the reason for the wrong sign of
$\bra{\JJ\cdot\BB}$ in the \FLASH code is its inability to reproduce the
relative strengths of small-scale and large-scale fields correctly.

\subsection{Total magnetic helicity production}

\begin{figure}[t!]\begin{center}
\includegraphics[width=\columnwidth]{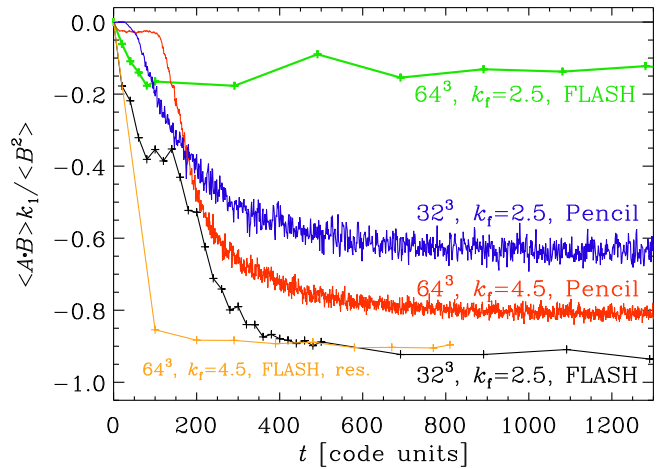}
\end{center}\caption[]{
Evolution of the fractional magnetic helicity for the case with $32^3$
mesh points, $\kf/k_1=2.5$, and $\eta^{\rm eff}=5\times10^{-5}$ (black
line), compared with the evolution in DNS with $32^3$ mesh points,
$\kf/k_1=2.5$, and $\eta=5\times10^{-5}$ (blue).
Also shown are a DNS with $64^3$ mesh points ($\kf/k_1=4.5$,
$\eta=5\times10^{-5}$, red line), and a solution with \FLASH with explicit
resistivity ($\kf/k_1=4.5$, $\eta=5\times10^{-5}$, orange line).
}\label{pabm}\end{figure}

An important question concerns the total magnetic helicity production
during the early small-scale and later large-scale dynamo processes.
We quantify this in terms of the evolution of the fractional magnetic
helicity defined as $\bra{\AAA\cdot\BB}k_1/\bra{\BB^2}$,
which is always between $+1$ and $-1$; see, e.g., \cite{KBTR10}.
Its evolution is shown in \Fig{pabm}, where we compare the results from
ideal simulations with those of DNS.
We find that both simulations produce negative magnetic helicity, but
the \FLASH code reaches about 90\%, while the expected value from the
DNS is only about 60\%.
By comparison, even with a larger scale separation of $\kf/k_1=4.5$
instead of $2.5$, we still only obtain about 80\% in the DNS.
This supports our earlier conclusion that the \FLASH code produces too
much power at large length scales.

We also see that, even at early times, the \FLASH code produces already
nearly 40\% magnetic helicity with $32^3$ mesh points and about 15\%
with $64^3$ mesh points.
The expected value based on the DNS is basically zero when $\kf/k_1=2.5$,
and about 2--3\% when $\kf/k_1=4.5$.
This difference at these early times is particularly remarkable, because
this is still the phase when the slow resistive evolution did not yet
have time to act.
It is even worse in the run with explicit magnetic diffusivity,
where a fractional helicity of 90\% is generated almost immediately.

\section{Discussion}

Our study has shown qualitative agreement between earlier resistive
simulations and the present ideal MHD simulations when both the resolution
is small ($32^3$ or $48^3$ mesh points) and the forcing wavenumber is small
($\kf/k_1=2.5$).
At higher resolution ($64^3$ mesh points), we find no large-scale dynamo
action at all (neither at $\kf/k_1=2.5$ nor at $4.5$).
It is curious, however, that the change between our $48^3$ and $64^3$
results is so abrupt.
Furthermore, the qualitatively different behavior in the form of
oscillations found by increasing $\kf/k_1$ from 2.5 to 4.5, is also
rather surprising.
In addition, as we just saw, the magnetic helicity is not really
zero in the $64^3$ simulation with $\kf/k_1=2.5$,
which is inconsistent with a solution to the truly ideal equations.
Thus, even though the absence of large-scale magnetic field generation
at late times in \Fig{pcomp} was compatible with an ideal evolution, the
moderate magnetic helicity production at early times in \Fig{pabm} is not.
Particularly worrisome is the case with explicit resistivity, which
always shows an effective resistivity that is several times larger than what is
put in, and there is rapid magnetic helicity production early on.

All these features -- the discontinuous dependence on resolution, the
oscillatory behavior in some cases, and the spurious magnetic helicity
production at early times -- suggest that the ideal state may
not be well defined and that different types of solutions may emerge
instead, at least in this specific case of an ideal MHD solver based
on the divergence-cleaning eight-wave scheme.
The behavior expected from the resistive evolution, as reproduced by
the {\sc Pencil Code} (\Figss{psat_pencil}{pabm}),
is not a typical outcome of ideal simulations,
except for some cases of low resolution, or with explicitly added
magnetic diffusivity.
How generic this departure from the resistive simulations is, however, remains open.
It would therefore be interesting to subject the problem discussed in
the present paper as a benchmark to other types of codes.
For codes that are kept numerically stable with some type of explicit
magnetic diffusion, e.g., through a modified scale dependence
such as magnetic hyperdiffusion, the final outcome can in principle be
predicted quantitatively, as was done by \cite{BS02}.
However, there could well be other schemes with quite different behaviors
that have not yet been anticipated.

In the eight-wave MHD solver invoked in \FLASHz, the constraint
$\nab\cdot\BB=0$ is solved through a divergence-cleaning algorithm
\citep{BB80}.
By calculating derivatives with a sixth order finite difference scheme,
we have verified that $\bra{(\nab\cdot\BB)^2}/\bra{\JJ^2}$ stays of the
order of $10^{-4}$, and does not increase.
In the {\sc Pencil Code}, by contrast, $\nab\cdot\BB=0$ is ensured by
solving directly for $\AAA$.
It might therefore be possible that the artificial magnetic
helicity production in \FLASH could be related to the use of the
divergence-cleaning algorithm.
This is not obvious, however, because the contribution from a gradient
correction to $\BB$ should not produce magnetic helicity if $\AAA$ is
computed in the Coulomb gauge.
In any case, as the resolution is increased from $48^3$ to $64^3$, not
only does the fractional helicity production during the non-resistive
phase decrease, but also the rate of magnetic helicity production
decreases.
This suggests that at sufficiently high numerical resolution, magnetic
helicity should be well conserved also in \FLASHz.
It would be interesting to see how magnetic helicity production is
affected by using instead the constrained transport algorithm
\citep{EH88}.

Triggered by the results reported in a preprint of the present
paper, Evan O'Connor (2019, private communication) examined our benchmark
problem with the {\tt SPARK} solver, which is planned to be part of
future releases of \FLASH (Sean Couch, 2019, private communication).
Even at low resolution of $16^3$ or $32^3$ mesh points, preliminary
results suggest a behavior that is similar to what is obtained
with the eight-wave solver at $64^3$ mesh points.
Details of this work will be reported elsewhere, but they do demonstrate
that there is no generic behavior of ideal MHD codes in general.
Based on the experience gathered so far, we can distinguish the following
types of behavior in the solution of a helical dynamo problem.
\vspace{-5pt}
\begin{itemize}\itemsep-2pt
\item[(i)]
{\em The behavior reproduced by using Euler potentials.}
In this case, no dynamo of any type has been reported as yet -- regardless
of the presence or absence of magnetic helicity \citep{Bra10}.
\item[(ii)]
{\em The ideal behavior reported in the present paper for $64^3$ mesh points
without explicit diffusivity.}
Small-scale dynamo action does occur, as expected, but there is no
large-scale dynamo action.
Such a behavior is in principle expected at infinite magnetic Reynolds number.
Magnetic helicity is produced at early times during the
kinematic phase (see \Fig{pabm}), and its amount can perhaps be estimated
in future based on the idea that a bihelical magnetic field is produced
with opposite signs at large and small length scales.
The latter contribution is thought to be absent, because the magnetic
field at small scales is not fully represented.
\item[(iii)]
{\em The resistive behavior of excessive amplitude reported by \cite{BS02}.}
Again, the small-scale dynamo is reproduced correctly, but the
large-scale dynamo is reproduced incorrectly in a predictable way: the
saturation amplitude of the large-scale magnetic field is too high by
a certain factor that can be computed a priori.
\item[(iv)]
{\em The resistive behavior found in the present work at low solution.}
Here, small-scale and large-scale dynamo action is possible.
The large-scale field evolves qualitatively as expected for a resistive
dynamo, but the details (saturation amplitude and the possibility of
oscillatory behavior) have not (yet) been possible to predict a priori.
\end{itemize}
The behavior (ii) may play an important role in future studies
of other dynamos (e.g., those with open boundary conditions and with
magnetic helicity fluxes).
This of course requires a better understanding of the quantitative
behavior of such a code, which will have to be addressed in future.

\section{Conclusions}

We have seen that, at low resolution, an ideal MHD code such as the
eight-wave scheme in \FLASH can reproduce certain aspects of resistive,
finite magnetic Reynolds number dynamos, although other aspects are still
not entirely physical.
For example, in a periodic system, the current helicity must approach
zero at late times, but no such tendency is found in the present
simulations (see \Fig{pjbm}).
Already at twice the resolution, however, the \FLASH code gives no
large-scale dynamo action at all.
This is, in principle, in agreement with the infinite magnetic Reynolds
number case, although the violation of magnetic helicity conservation
at early times speaks against this.
Real systems, on the other hand, are not fully homogeneous and cannot
be described by periodic boundary conditions.
This can lead to the occurrence of magnetic helicity fluxes \citep{BF00}.

It would in future be interesting to extend our studies to systems that
do possess a magnetic helicity flux \citep{HB10,MCCTB10,DSGB13,Bra18}.
In view of our results, however, we cannot take it for granted that the
magnetic field evolution in poorly resolved systems reproduces in any
way the behavior expected for a standard Spitzer resistivity.

Of course, modern simulations tend to have a numerical resolutions much
larger than $32^3$, but at the same time, one usually captures much more
complex physical processes covering a large range of length scales.
At the smallest scale, therefore, the effective resolution is again
just barely enough to resolve the details of magnetic structures.
In this sense, our work has implications for the study of dynamos with
ideal codes at any resolution.
It remains therefore mandatory to subject any dynamo simulation to a
proper convergence test with fixed explicit resistivity.

In this paper, we have focused on magnetic helicity produced or dissipated
in a dynamo setup where kinetic helicity is constantly being supplied.
Our findings are not, however, restricted to dynamo applications,
and certainly not to large-scale dynamos, because we have demonstrated
that magnetic helicity can be generated even at very short times before
the large-scale dynamo can have acted.
Another class of applications would be that of decaying MHD turbulence,
which has been studied extensively by many groups both without magnetic
helicity \citep{MLKB98,Zra14,BKT15} and with magnetic helicity
\citep{CHB01,BJ04,BK17} using both ideal MHD codes such as {\tt ZEUS}
and non-ideal ones such as the {\sc Pencil Code}.
The behaviors found with the different codes are similar in that the
inverse magnetic transfer is being reproduced in both cases, except that
{\tt ZEUS} showed more resistive behavior \citep{Reppin}.
This became strikingly clear in simulations of the inverse magnetic
transfer found in the nonhelical case \citep{BL14}.
Further studies would be of interest, especially at large numerical
resolutions of the order of $2304^3$ mesh points, as already
done with the {\sc Pencil Code} \citep{BKT15}.

\acknowledgments
We thank Matthias Rheinhardt for useful comments on the manuscript and
Evan O'Connor for sharing with us the results of his simulations with \FLASHz.
We are grateful for the suggestions made by an anonymous referee.
This work was performed at the Aspen Center for Physics, which is
supported by National Science Foundation grant PHY-1607611.
We enjoyed the stimulating atmosphere during the Aspen program on
the Turbulent Life of Cosmic Baryons.
We also acknowledge the organizers and participants of the
Nordita-supported program on Solar Helicities in Theory and Observations
for a lively discussion on diffusionless dynamos.
This research was supported in part by the Astronomy and Astrophysics
Grants Program of the National Science Foundation (grants 1615100 and 1715876).
We acknowledge the allocation of computing resources provided by the
Swedish National Allocations Committee at the Center for Parallel
Computers at the Royal Institute of Technology in Stockholm.

\appendix
\section{Late saturation phase}
\label{Late}

To understand the origin of \EEq{EEMsat}, we use \Eq{dABdt}, introduce
mean fields, $\meanBB$, as suitably defined planar averages, and
define fluctuations correspondingly as $\bb=\BB-\meanBB$, and likewise
for the magnetic vector potential $\aaaa=\AAA-\meanAA$ and the magnetic
current density $\jj=\JJ-\meanJJ$, respectively.
\EEq{dABdt} then becomes
\begin{equation}
{\dd\over\dd t}\bra{\meanAA\cdot\meanBB}=
-2\eta\bra{\meanJJ\cdot\meanBB}-2\eta\bra{\jj\cdot\bb},
\label{dAmBmdt}
\end{equation}
where we have ignored the time derivative of $\bra{\aaaa\cdot\bb}$,
because the small-scale magnetic field has saturated at $t=t_{\rm sat}$
(see \Fig{pcomp_early}) and is approximately constant during the late
saturation phase, $t>t_{\rm sat}$.
Next, we approximate $\bra{\meanAA\cdot\meanBB}\approx-\bra{\meanBB^2}/k_1$,
$\bra{\meanJJ\cdot\meanBB}\approx-\bra{\meanBB^2}k_1$, and
$\bra{\jj\cdot\bb}\approx+\bra{\bb^2}\kfeff$.
Finally, we approximate $\bra{\bb^2}/2\approx\EEK$, and obtain
\begin{equation}
\left(2\eta k_1^2+{\dd\over\dd t}\right)\frac{\bra{\meanBB^2}}{2}=2\eta k_1\kfeff\EEK,
\end{equation}
which can be integrated to yield \Eq{EEMsat}, where
$\EEM-\EEK=\bra{\meanBB^2}/2$ has been used.


\end{document}